\documentclass[10pt,twocolumn,superscriptaddress,prl,showpacs,floatfix,aps]{revtex4-2}
\usepackage[utf8]{inputenc}
\usepackage{amsmath}
\usepackage{amssymb}
\usepackage{graphicx}
\usepackage{svg}
\usepackage{xspace}
\usepackage{mathtools}
\usepackage{amsfonts}
\usepackage{bm}
\usepackage{bbm}
\usepackage{dsfont}
\usepackage{soul}
\usepackage{comment}
\usepackage{physics}
\usepackage[normalem]{ulem}
\usepackage{babel}
\usepackage{xcolor}
\usepackage[final]{hyperref} % adds hyper links inside the generated pdf file
\hypersetup{
	colorlinks=true,    % false: boxed links; true: colored links
	linkcolor=blue,    % color of internal links
	citecolor=blue,    % color of links to bibliography
	filecolor=magenta,   % color of file links
	urlcolor=blue     
}
\usepackage[american,siunitx]{circuitikz}
\usepackage{amsmath}
\usepackage{orcidlink}
\usetikzlibrary{shapes,decorations.pathmorphing}
\newcommand{\ba}{\begin{array}}
\newcommand{\ea}{\end{array}}
\newcommand{\be}{\begin{equation}}
\newcommand{\ee}{\end{equation}}
\newcommand{\bea}{\begin{eqnarray}}
\newcommand{\eea}{\end{eqnarray}}
\newcommand{\Diss}{\mathcal D}
\newcommand{\Li}{\mathcal L}
\newcommand{\rss}{\rho^{\mathrm{st}}}
\newcommand{\nrm}[1]{\left\lVert#1\right\rVert}
\newcommand{\td}{\mathrm{TD}}
\newcommand{\rp}{\mathrm{RP}}
\newcommand{\id}{\mathds{1}}
\newcommand{\FDR}{\operatorname{FDR}}

\begin{document}

\title{Predicting Quantum Mpemba Effects from Quantum Speed Limits}
 
\author{Andrea Nava~\orcidlink{0000-0002-6729-3800}}
\affiliation{Institut f\"ur Theoretische Physik, Heinrich-Heine-Universit\"at, D-40225  D\"usseldorf, Germany}

\author{Rosario Fazio~\orcidlink{0000-0002-7793-179X}}
\affiliation{The Abdus Salam International Center for Theoretical Physics, Strada Costiera 11, I-34151 Trieste, Italy}
\affiliation{
Dipartimento di Fisica, Università di Napoli ``Federico II'', Monte S. Angelo, I-80126 Napoli, Italy}

\author{Reinhold Egger~\orcidlink{0000-0001-5451-1883}}
\affiliation{Institut f\"ur Theoretische Physik, Heinrich-Heine-Universit\"at, D-40225  D\"usseldorf, Germany}

\begin{abstract}
 Can anomalous Quantum Mpemba Effects be anticipated before a system comes close to its stationary state? 
 We employ Quantum Speed Limits as geometric indicators of Quantum Mpemba Effects, arguing that Mpemba physics is already imprinted in speed limits by benchmarking dissipative qubits and open spin chains. Our results show that short-time speed limits can predict Quantum Mpemba Effects at low cost and with reasonable precision over broad parameter regimes. This predictive power implies a useful application of speed limits for open many-body quantum systems.
\end{abstract}

\maketitle

\emph{Introduction.---}The Quantum Mpemba Effect (QME) challenges the expectation that an initial state farther away from the final stationary state takes longer to relax; see Refs.~\cite{Ares_2025,Teza_2025} for reviews. 
QMEs exist both for unitary and open quantum systems; we focus on the open case.  In Markovian dynamics, anomalous relaxation can arise when the initially more distant state has a smaller projection onto the slow modes of the time evolution generator ${\cal L}$ of the system \cite{Lu_2017,Carollo_2021,Nava_2024}; eliminating the slowest mode from the initial state even accelerates the asymptotic decay rate. While such spectral mechanisms explain late-time relaxation, finding favorable initial conditions requires to resolve the slow subspace of ${\cal L}$ or to propagate many candidate states. 
Especially when scanning broad parameter regions in many-body quantum systems, such an endeavor becomes highly challenging and costly. This observation motivates the central question investigated in this Letter: Can features of the initial transient dynamics already predict QMEs with reasonably good accuracy?  This naturally raises the question if there are signatures of QMEs imprinted into Quantum Speed Limits (QSLs) \cite{Mandelstam1991,Margolus_1998,delCampo_2013,Taddei_2013,Deffner_2013,Pires_2016,Deffner_2017,Deffner_2017,Funo_2019,Campaioli2019,Pati_2025,Srivastav_2025}, which impose lower bounds on the time  required for a quantum system to evolve from an initial to a target state.   
 In open systems, QSLs not only represent fundamental lower bounds on relaxation times but are also useful for different applications, e.g., quantum control \cite{Caneva_2009,Cao_2024,Basilewitsch_2024}, sensing \cite{Herb_2024}, quantum batteries \cite{Gyhm_2024,Shrimali_2024}, or many-body scrambling \cite{Vikram2024}. We here show that QSLs can have low-cost predictive power for QMEs, which in turn will drastically simplify the search for parameter regimes exhibiting QMEs in different platforms, including many-body systems.

In general, QSLs relate a certain distance in state space ($d$) to an averaged velocity ($\bar v$), where the true relaxation time $\tau$ is bounded from below, $\tau\ge \tau^{\rm QSL}=d/\bar v$. Clearly, a lower bound need not provide an accurate estimate for $\tau$, and comparing two lower bounds need not reproduce the ordering of actual times. Nevertheless, the geometric quantities entering a QSL can retain crucial information relevant to that ordering. In fact, trajectory geometry has been connected to Mpemba effects \cite{Srivastav_2025,Qian_2025,Hong2026}, and initial-time complexity diagnostics can be related to Mpemba-like inversions \cite{Beetar_2026}. 
Focusing on Markovian systems  described by Lindblad equations \cite{breuer2007theory}, we show that QSLs directly connect short-time geometry and QMEs.
As measure for the state distance to stationarity, we employ the trace distance (TD). Using QSL-based scores based on either TD or relative purity (RP),
we examine the predictive power of short-time geometry for QMEs. For given ${\cal L}$, we use a common observation time $t^\ast<\tau$ for the studied set of initial states.
This choice permits a direct distinction between information available before $t^\ast$ and information obtained by completing a trajectory. 
We provide analytical derivations of short-time geometric QSL scores, compare them to numerical results for their finite-time extensions, and benchmark their performance in dissipative qubit and many-body spin-chain systems. 
Based on these results, we show that in systems with a small Liouvillian gap and/or early last-time crossings, QSL-based short-time scores already yield reliable and low-cost predictions for QMEs, even for many-body systems. 

\emph{Lindblad dynamics and QSLs.---}We consider a finite-dimensional Lindblad equation \cite{breuer2007theory},
\begin{equation}\label{eq:lindblad}
\dot\rho=\Li\rho=-i[H,\rho]+\sum_\mu\Diss[L_\mu]\rho,
\end{equation}
governing the dynamics of the system state $\rho(t)$, with $\hbar=1$ and $\Diss[L]\rho=L\rho L^\dagger-\tfrac12\{L^\dagger L,\rho\}$.
The Liouvillian $\Li$ contains a unitary piece due to a Hamiltonian $H$ and dissipative terms encoded by a set of jump operators $\{L_\mu\}$.
Starting from initial state $\rho(0)=\rho_0$, we assume convergence of $\rho(t)$ to a stationary state $\rss$ at $t\to \infty$ as measured by the trace distance $D(t)=\tfrac12\nrm{\rho(t)-\rss}_1$, with the 1-norm
$\nrm{\cdot}_1$.
Since $\rss$ is approached only for $t\to \infty$, we define a threshold value $0<\epsilon\ll 1$ such that the relaxation time is $\tau=\inf\{t\geq0:D(t)\leq\epsilon\}$.  For two states $\rho_A(t)$ and $\rho_B(t)$ satisfying the initial ordering $D_A(0)>D_B(0)>\epsilon$, the QME occurs if $\tau_A<\tau_B$ \cite{Teza_2025}. 

For given initial state and state paths observed up to time $t^\ast\le\tau$, we define the QSL indicators  \cite{Rosal2025}
\begin{align}
\tau^{\rm QSL}_{\td}(t^\ast)&= \frac{\nrm{\rho(t^\ast)-\rho_0}_1}
{\int_0^{t^\ast}\!\frac{dt}{t^\ast}\,\nrm{\Li\rho(t)}_1},\label{eq:qtd}\\
\tau^{\rm QSL}_{\rp}(t^\ast)&= \frac{|\Tr\{\rho_0[\rho(t^\ast)-\rho_0]\}|}
{\nrm{\rho_0}_2\,\int_0^{t^\ast}\!\frac{dt}{t^\ast}\,\nrm{\Li\rho(t)}_2},
\label{eq:qrp}
\end{align}
with the Schatten $k$-norm $\nrm{X}_k=({\rm Tr}|X|^k)^{1/k}$. 
Here $\tau^{\rm QSL}_{\td}(t^\ast)$ employs a TD measure taken with respect to the initial state,
while $\tau^{\rm QSL}_{\rp}(t^\ast)$ is instead based on the RP.
Applying the integral triangle inequality and the Hilbert-Schmidt Cauchy-Schwarz inequality, see the End Matter, we obtain the QSL bounds 
\begin{equation}\label{QSLbound}
0\leq \tau^{\rm QSL}_{j\in \{\td,\rp\}}(t^\ast)\leq t^\ast\leq \tau.
\end{equation}  
In the context of QMEs, and in contrast to other QSL applications, we typically consider short times  ($t^\ast\ll \tau$).
While actual relaxation times are poorly represented by short-time QSL bounds, they can be very useful for predicting Mpemba orderings of initial-state pairs.
Below we often keep the index $j$ implicit.

\emph{Diagnostics.---}Assuming the initial ordering  $D_{A}(0)>D_{B}(0)>\epsilon$, we use $\tau^{\rm QSL}_{A}(t^\ast)<\tau^{\rm QSL}_{B}(t^\ast)$ as QSL prediction of a QME. While the QSL bounds \eqref{QSLbound} are rigorous, the inference about Mpemba ordering is an empirical diagnostic. For benchmarking,  we restrict the eligible pair populations to states satisfying
\begin{equation}\label{eq:eligible}
{\cal E}_{t^\ast}=\{(A,B):{} D_{A}(0)>D_{B}(0)>\epsilon,\, {} t^\ast<\min(\tau_{A},\tau_{B})\}.
\end{equation}
A true Mpemba inversion $\tau_{A}<\tau_{B}$ is then denoted by $M$, whereas $\widehat M$ corresponds to the QSL prediction.
The prevalence and prediction coverage of Mpemba pairs on ${\cal E}_{t^\ast}$ are respectively defined as
\begin{equation}\label{pi0q}
    \pi_0 =\Pr(M\mid\mathcal E_{t^\ast}),\quad q =\Pr ( \widehat{M}\mid\mathcal E_{t^\ast} ),
\end{equation}
where the $t^\ast$-dependence is kept implicit.
We compute the conditional probabilities \cite{Powers2011,Saito2015}
\begin{equation}\label{eq:metrics}
P=\Pr(\widehat M\mid M,{\cal E}_{t^\ast}),\quad
\FDR=\Pr(\neg M\mid\widehat M,{\cal E}_{t^\ast}),
\end{equation}
where $P$ is the  recall  (\emph{aka} predictive power) and $\FDR$ the false discovery rate.  Another useful diagnostic is precision enrichment,
\begin{equation}\label{precisionenrichment}
{\rm PE} = \frac{1-\FDR}{\pi_0} = \frac{P}{q}.
\end{equation}
We compare actual data for $(P,\FDR,{\rm PE})$ to a random baseline, 
\begin{equation}\label{eq:random}
(P,\FDR,{\rm PE})_{\rm rnd}=(q, 1-\pi_0, 1), 
\end{equation}
where each eligible pair is flagged with probability $q$ independently of its true label. 
In general, there are no universally satisfactory criteria for $(P,\FDR,{\rm PE})$ \cite{Powers2011,Saito2015}. However, 
in practice, predictors are deemed useful if they give substantial precision enrichment, ${\rm PE}>1$, together with large recall $P$ at computationally low cost.
For additional details on benchmarking, see the End Matter.

We next turn to the case of a dissipative qubit, where we benchmark short-time QSL predictions in detail. 
Equipped with these insights, we then move on to many-body systems and study open spin-chain models.

\emph{Local geometry of the predictor.---}We first consider a dissipative qubit described by Eq.~\eqref{eq:lindblad}, where analytical progress is possible. Writing $\rho(t)=\frac12[\id+\mathbf r(t)\cdot\bm\sigma]$ with the Bloch vector $\mathbf r(t)$ and the 
vector of Pauli matrices $\bm \sigma$, for an initial state with $\mathbf r(0)=\mathbf r_0$, 
the Lindblad equation yields an affine equation for the Bloch vector,
\begin{equation}\label{affine}
\dot{\mathbf r}=\Lambda\mathbf r+\mathbf c,
\quad \mathbf v_0=\Lambda\mathbf r_0+\mathbf c,
\quad \mathbf a_0=\Lambda\mathbf v_0,
\end{equation}
where the $3\times 3$ matrix $\Lambda$ and the vector $\mathbf c$ depend on $H$ and on the jump operators $L_\mu$ \cite{breuer2007theory,Nava_2025}.  In Eq.~\eqref{affine}, we also give the initial velocity $\dot{\mathbf r}(0)=\mathbf v_0$ and acceleration $\ddot{\mathbf r}(0)=\mathbf a_0$.
The TD-QSL indicator \eqref{eq:qtd} then yields $\tau^{\rm QSL}_{\td}=t^\ast C/L$ with the chord ($C$) and arc ($L$) length of the Bloch trajectory \cite{Rosal2025}. For $\mathbf v_0\ne 0$, short-time expansion in $t^\ast$ obtains 
\begin{equation}
\tau^{\rm QSL}_{\td}(t^\ast)=t^\ast-\frac{k(\mathbf r_0)}{24}t^{\ast\, 3}+O(t^{\ast \,4}),
\label{eq:shorttd}
\end{equation}
with the squared initial turning rate 
\begin{equation}
k(\mathbf r_0)=\frac{|\mathbf v_0|^2|\mathbf a_0|^2-(\mathbf v_0\cdot\mathbf a_0)^2}{|\mathbf v_0|^4}
=\left|\left.\frac{d\widehat{\mathbf v}}{dt}\right|_0\right|^2.
\label{eq:turn}
\end{equation}
 Consequently, this QSL score does not simply correspond to $-\dot{D}(0)$.
Similarly, the RP construction \eqref{eq:qrp} yields the short-time expansion
\begin{equation}
\tau^{\rm QSL}_{\rp}(t^\ast)=b(\mathbf r_0)t^\ast+O(t^{\ast\,2}),\quad
b(\mathbf r_0)=\frac{|\mathbf r_0\cdot\mathbf v_0|}
{\sqrt{1+|\mathbf r_0|^2}\,|\mathbf v_0|},
\label{eq:shortrp}
\end{equation}
which depends on the initial alignment angle between $\mathbf{r}_0$ and $\mathbf{v}_0$ and on the Bloch-vector length. 
In general, $k\ge 0$ and $0\le b\le 1$. 

QSL rankings can be obtained by computing the short-time scores $s_{\td}=-k$ and $s_{\rp}=b$ for different $\rho_0$, with smaller scores predicting faster relaxation. 
These scores follow analytically from ${\cal L}$ and $\rho_0$, without need for dynamical propagation, where TD- vs RP-based scores retain distinct geometric information. 
The RP can be unchanged even for different states, with $b=0$ unrelated to proximity to stationarity. The TD construction avoids that particular artifact, but instead remains insensitive, e.g., to radial amplitudes. Indeed, for the initial-state class $\rho_{0,\tilde c}=\rss+ \tilde c X$, with $\tilde c>0$ and  traceless Hermitian  $X$, both the numerator and the denominator in Eq.~\eqref{eq:qtd} scale with $\tilde c$. Hence $\tau^{\rm QSL}_{\td}(t^\ast)$ becomes independent of $\tilde c$, in contrast to the full relaxation time $\tau$. Short-time geometry based on a single QSL score therefore cannot properly rank all relaxation times. However, QSL predictions obtained from different measures can give complementary insights.

\begin{figure}
\centering
\includegraphics[width=\linewidth]{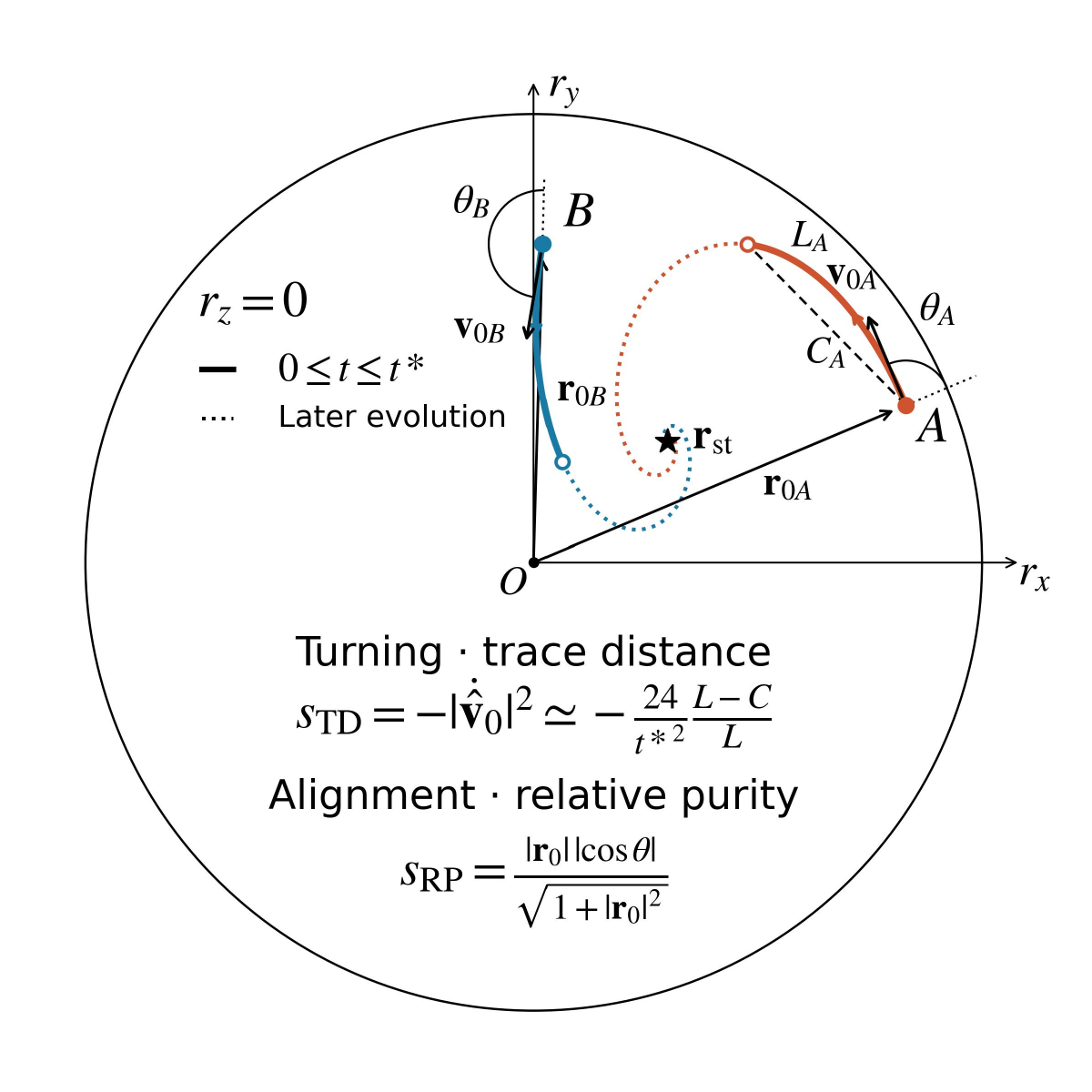}
\caption{Short-time QSL geometry in the invariant Bloch plane $r_z=0$ for the planar qubit model in Eq.~\eqref{eq:A} with $H=\sigma_z$ and $(\gamma,\eta)=(0.45,0.33)$, using $t^\ast=0.5$ and $\epsilon=10^{-2}$. For initial states with ${\bf r}_{0 A}=(0.83,0.35)$ (red) and ${\bf r}_{0 B}=(0.02,0.71)$ (blue), bold segments show ${\bf r}(t)$ for $0\le t\le t^\ast$; dotted curves show the later relaxation toward ${\bf r}_{\rm st}$ parametrizing $\rss$. Open (filled) circles mark $t^\ast$ ($t=0$), and the star marks ${\bf r}_{\rm st}$. The QSL indicator \eqref{eq:qtd} equals $t^\ast C/L$ with the chord ($C$) and arc ($L)$ length. Short-time QSL scores are $s_\td=-k$ and $s_\rp=b$.}
\label{fig1}
\end{figure}

To separate coherent rotation from anisotropic damping effects, we study an example with  $H=\sigma_z$ and a single jump operator $L_1=\sqrt{\gamma}\,\left(\sigma_x+\sigma_y+i\eta\sigma_z\right)$, with damping strength $\gamma>0$ and real anisotropy parameter $\eta$. The equatorial Bloch plane $r_z=0$ is then invariant, 
with its affine dynamics \eqref{affine} described by
\begin{equation}
\Lambda_{\perp}=\begin{pmatrix}
-2\gamma(1+\eta^2)&2(\gamma-1)\\
2(\gamma+1)&-2\gamma(1+\eta^2)\end{pmatrix},
\quad
\mathbf c_{\perp}=\begin{pmatrix}4\gamma\eta\\-4\gamma\eta\end{pmatrix}.
\label{eq:A} 
\end{equation}
The planar eigenvalues are $\lambda_\pm=-2\gamma(1+\eta^2)\pm2\sqrt{\gamma^2-1}$, while the unexcited $z$ component decays at rate $4\gamma$. For $\gamma$ below the exceptional point $\gamma=1$, the planar modes form a complex pair, while above it, they imply real-valued decay. The above model is analytically solvable, see the 
End Matter.  The geometric information entering both QSL predictors is illustrated in Fig.~\ref{fig1} using trajectories of Eq.~\eqref{eq:A}. The shown initial-state pair $(A,B)$ exhibits the QME for $\epsilon=10^{-2}$: We find $D_A(0)>D_B(0)>\epsilon$ but $\tau_A<\tau_B$ with $t^\ast<{\rm min}(\tau_A,\tau_B)$. This QME is correctly ranked by both scores: $s_A<s_B$ holds for $s_\td=-k$ and for $s_\rp=b$. 

\begin{figure}[t]
\centering
\includegraphics[width=\linewidth]{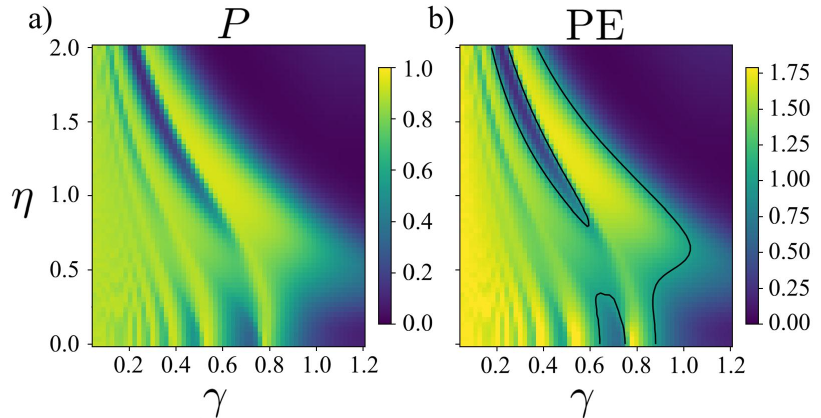}
\caption{Color-scale plots in the $\gamma$--$\eta$ plane of the planar qubit model \eqref{eq:A} for (a) the recall $P$ and (b) the precision enrichment ${\rm PE}$,  see Eqs.~\eqref{eq:metrics} and \eqref{precisionenrichment}, respectively. We use the short-time QSL score $s_{\rm TD}=-k$ with $\epsilon=10^{-2}$, averaging over all distinct pairs formed from 600 quasi-uniformly distributed initial Bloch-disk states with equal-area radial spacing, golden-angle azimuths, and antipodal pairing. Black curves in (b) indicate ${\rm PE}_{\rm rnd}=1$, see Eq.~\eqref{eq:random}.}
\label{fig2}
\end{figure}

In Fig.~\ref{fig2}, we show the recall ($P$) and the precision enrichment (PE) predicted by $s_\td=-k$ for the model in Eq.~\eqref{eq:A} as color-scale plots in the $\gamma$--$\eta$ plane. We compute these diagnostic quantities 
by counting the number of successful QME predictions for all initial-state pairs taken from a set of
600 quasi-uniformly distributed Bloch-disk states. The  QSL score $s_{\rm RP}=b$ yields a similar picture, see the End Matter, where we also provide results for the FDR, $\pi_0$, and $q$, see Eqs.~\eqref{pi0q} and \eqref{eq:metrics}. 
We observe from Fig.~\ref{fig2} that for weak damping, QMEs are predicted over wide parameter regions, with ${\rm PE}>1$ and high recall $P$. 
At the same time, the FDR can be rather large because of the small underlying prevalence $\pi_0$. For $\gamma>1$, the score $s_\td=-k$ can reach zero and therefore fail, see the End Matter. 
Nonetheless, Fig.~\ref{fig2} shows that once a point where the score has predictive power has been identified, 
a finite region around it is also described by it, indicating robustness against small parameter changes. 
We note that parameter regions where the QSL scores predict a QME are often characterized by early-time crossings of the TD curves. Such regions are of interest since 
they may cause rather large Mpemba speedups \cite{Teza_2025}.    

\begin{figure}[t]
\includegraphics[width=\linewidth]{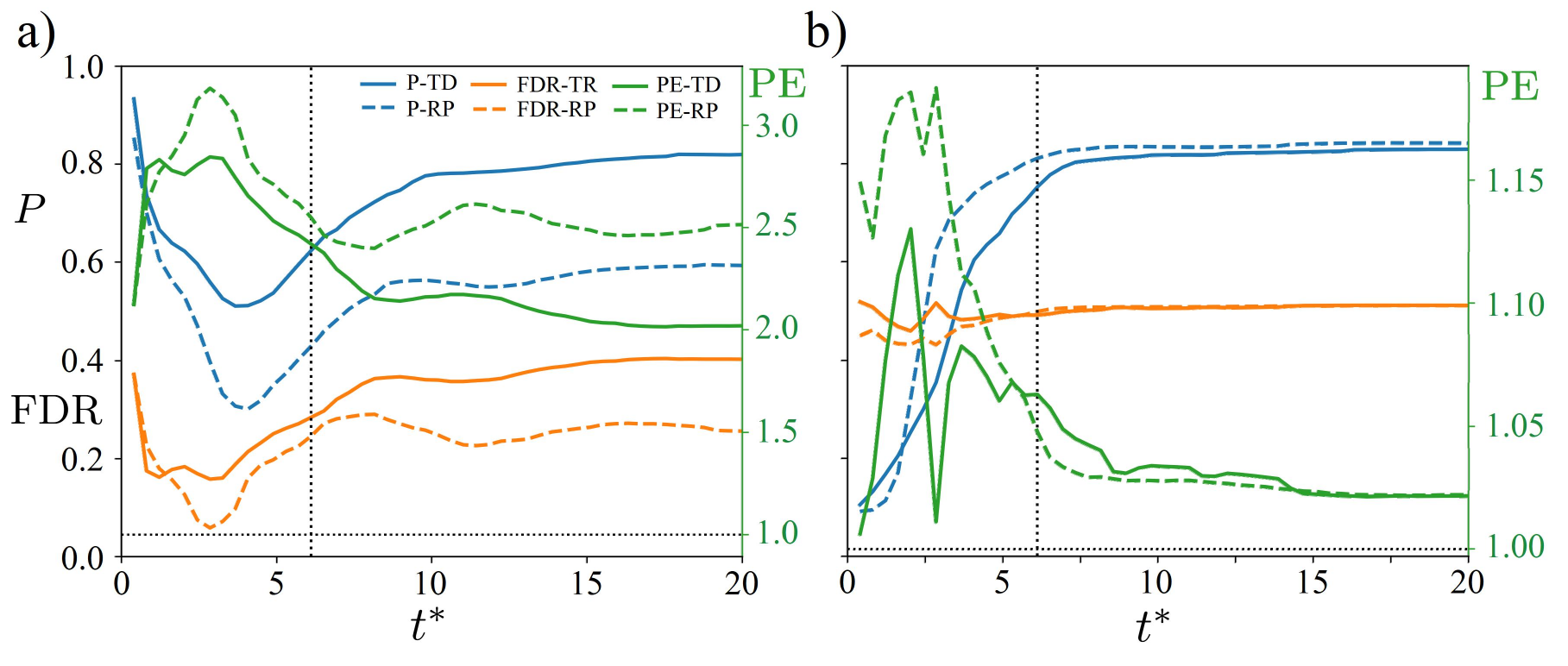}
\caption{Diagnostic measures ($P$, PE, FDR) vs $t^\ast$ for the collective-spin model \eqref{eq:collective} 
with $S=10$, $\omega_0=1$, $(\kappa_-,\kappa_+,\kappa_z)=(1,0.3,0.5)$, and $\epsilon=10^{-3}$, using the TD-based (solid) and RP-based (dashed curves) QSL indicators \eqref{eq:qtd} and \eqref{eq:qrp}, respectively.   The dashed vertical line corresponds to the inverse gap of ${\cal L}$, the dashed horizontal line indicates ${\rm PE}_{\rm rnd}=1$. (a) 
Results averaged over the initial-state class  $|\psi_1\rangle$ in Eq.~\eqref{eq:three_dicke_family}.  
(b) Same as in panel (a) but for the initial-state class $|\psi_2\rangle$ in Eq.~\eqref{eq:dicke_initial_state}. 
}
\label{fig3}
\end{figure}

\emph{Many-body systems.---}We now move on to many-body quantum systems, starting with a collective-spin model related to the boundary time crystal studied, e.g., in Ref.~\cite{Iemini_2018}.  With total spin $S$, the Liouvillian is
given by
\begin{equation}\label{eq:collective}
\Li\rho=-i[\omega_0S^x,\rho]+ \frac{1}{S}\sum_{a\in\{+,-,z\}}\kappa_a\Diss[S^a]\rho,
\end{equation}
where the collective spin operators $(S^x,S^y,S^z)$ with $S^\pm=S^x\pm i S^y$ preserve the maximal-spin ($S$) sector.  
As initial states, we consider either superposition states $|\psi_1\rangle$ of three consecutive Dicke states parametrized by angles $(\theta,\phi)$,
\begin{equation}\label{eq:three_dicke_family}
|\psi_1(\theta,\phi)\rangle=\cos\theta|S,-1\rangle+\sin\theta\cos\phi|S,0\rangle+\sin\theta\sin\phi|S,1\rangle,
\end{equation}
or alternatively, with $m(\phi)=\mathrm{nint}(S\cos\phi)$ and the nearest-integer function nint,
\begin{equation} \label{eq:dicke_initial_state}
|\psi_2(\theta,\phi)\rangle =
\cos\left(\frac{\theta}{2}\right)|S,S\rangle
+\sin\left(\frac{\theta}{2}\right)|S,m(\phi)\rangle,
\end{equation}
up to a normalization factor for $m(\phi)=S$.
The $t^\ast$-dependence of the diagnostic measures $P$, PE, and FDR is shown in Fig.~\ref{fig3} for the TD- and RP-based QSL indicators \eqref{eq:qtd} and \eqref{eq:qrp}.  Here we
have averaged over all distinct pairs taken from 625 initial states with quasi-uniformly distributed angles $(\theta,\phi)$  over a $25\times 25$ grid with $\theta \in [0,\pi]$ and $\phi \in [0,2\pi)$. 
In Fig.~\ref{fig3}(a), we study the initial states $|\psi_1\rangle$ in 
Eq.~\eqref{eq:three_dicke_family}, where a histogram analysis 
reveals a dominance of early last-time crossings of TD curves, see the End Matter.
In Fig.~\ref{fig3}(b), we instead consider the initial states $|\psi_2\rangle$ in Eq.~\eqref{eq:dicke_initial_state}, where late-time crossings prevail.  In Fig.~\ref{fig3}(a), we accordingly observe that excellent predictions follow from short-time QSL indicators, with $P\approx 0.8$,  ${\rm PE}\approx 3$, and FDR $\alt 0.2$.  Both TD and RP scores give qualitatively similar results. In contrast, in Fig.~\ref{fig3}(b), while the FDR is only weakly
$t^\ast$-dependent,  $P(t^\ast)$ becomes large only for $t^\ast$ above the inverse gap of ${\cal L}$
and the PE is smaller, even though ${\rm PE}>1$ for all $t^\ast$.  
If late-time crossings proliferate, QSL predictions are thus more costly and also less precise. 
We note that the sharp features 
observed at early times for ${\rm PE}(t^\ast)$ in Fig.~\ref{fig3}(b) are due to spectral changes affecting the trace norms and near-tie cases.

\begin{figure}[t]
\includegraphics[width=1.0\linewidth]{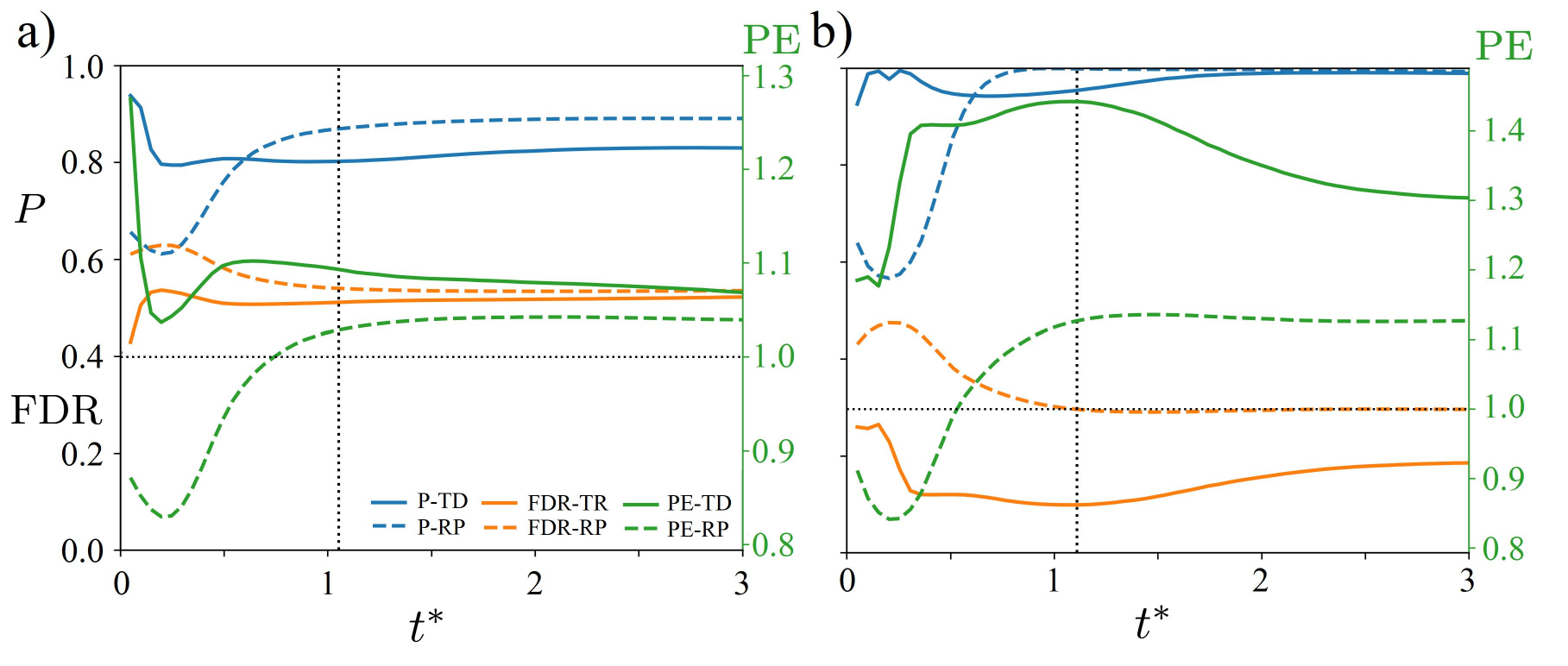}
\caption{Diagnostic measures ($P$, PE, FDR) vs $t^\ast$ for the open spin chain model \eqref{eq:chain} with  $N=6$, $h_{x,j}=1$, $(\gamma_-,\gamma_+,\gamma_z)=(1,0.3,0.5)$, and $\epsilon=10^{-3}$. Results are shown for the TD-based (solid) and RP-based (dashed curves) QSL indicators, averaging over pairs drawn from the initial states \eqref{eq:spin_chain_family}. Dashed vertical lines indicate the inverse gap of ${\cal L}$, and the horizontal dashed line gives PE$_{\rm rnd}=1$.  (a) Interacting case with $h_{z,j}=0.20+0.06(j-1)$, $J_x=J_y=0.4,$ and $J_z=0.8$. (b) Noninteracting case with $h_{z,j}=J_{\alpha}=0$.   
}
\label{fig4}
\end{figure}

To go beyond the Dicke manifold, we next study an open spin-$1/2$ chain with $N$ sites and open boundary conditions described by 
\begin{equation}
H=\frac12\sum_{j=1}^{N}(h_{x,j}\sigma_j^x+h_{z,j}\sigma_j^z)
+\sum_{j=1}^{N-1}\sum_{\alpha=x,y,z}\frac{J_\alpha}{4}\sigma_j^\alpha\sigma_{j+1}^\alpha,
\label{eq:chain}
\end{equation}
where the local jump operators $L_{j,a\in\{-,+,z\}}= \sqrt{\gamma_a} \sigma_j^a$ with $\sigma_j^\pm=(\sigma_j^x\pm i\sigma_j^y)/2$
permit transitions between total-spin sectors. As initial states, we consider the staggered product states
\begin{equation} \label{eq:spin_chain_family}
|\psi(\theta,\phi)\rangle=\bigl[|u(\theta,\phi)\rangle\otimes|u(\pi-\theta,-\phi)\rangle\bigr]^{\otimes3} 
\end{equation}
with $|u(\theta,\phi)\rangle=\cos(\theta/2)|0\rangle+e^{i\phi}\sin(\theta/2)|1\rangle$, where we take 625 quasi-uniformly distributed values of $(\theta,\phi)$ on a $25\times25$ grid, with $\theta \in [0,\pi]$ and $\phi \in [0,2\pi)$. 
Results are then averaged over all distinct pairs of these initial states. For the interacting case  in Fig.~\ref{fig4}(a), the TD-QSL indicator predicts high recall $P\simeq0.8$, ${\rm PE}>1$, and $\FDR\simeq0.5$. 
A short-time TD-based QSL score here works well, while an RP-based indicator should be evaluated at later
times $t^\ast$.  For the noninteracting case, see Fig.~\ref{fig4}(b) with $h_{z,j}=J_\alpha=0$, even higher recall
and smaller FDR are found. Now deviations between TD- and RP-based indicators are larger, but both give ${\rm PE}>1$ at large $t^\ast$. These examples demonstrate the usefulness and limitations of QSL scores.

\emph{Many-body short-time QSL scores.---}The qubit scores $s_\td=-k$ and $s_\rp=b$ can be extended to many-body systems. Let us define traceless Hermitian operators for the initial velocity, $V=\dot\rho_0=\Li\rho_0$, and  the initial acceleration, $A=\ddot\rho_0=\Li^2\rho_0$.
Assuming that $V$ is nonsingular, Eq.~\eqref{eq:qtd} yields  
$\tau_{\td}^{{\rm QSL}}(t^\ast) =t^\ast-\frac{K}{24}t^{\ast\, 3}+O(t^{\ast\,4}),$
where $K=\frac{\dd^2F_V[A,A]}{F(V)}$ with $F(X)=\|X\|_1$. 
Here the Hessian evaluated twice on $A$ is the directional second derivative,
$\dd^2F_V[A,A]=\left.\frac{\dd^2}{\dd s^2}\|V+sA\|_1\right|_{s=0}.$
Thus the short-time QSL score is $s_{\td}=-K$, which is determined by $\Li\rho_0$ and $\Li^2\rho_0$.
The expansion is not necessarily uniform in system size or in the smallest absolute eigenvalue of $V$.  
Convexity of the trace norm implies $K\ge 0$.  
A spectral expression for $K$ follows by diagonalizing the initial velocity, 
$V=\sum_i v_i|i\rangle\langle i|$ with $v_i\ne 0$.  With $a_{ij}=\langle i|A|j\rangle$, one finds \cite{Lewis2001} 
\begin{equation}
K=\frac{4}{\|V\|_1}
\sum_{\substack{i:\, v_i>0\\j:\, v_j<0}}
\frac{|a_{ij}|^2}{v_i+|v_j|},
\label{eq:spectral}
\end{equation}
which quantifies the mixing of positive and negative spectral subspaces of $V$ due to acceleration; this is a norm-dependent bending measure. Similarly, for the RP-based indicator \eqref{eq:qrp}, $\tau_{\rp}^{\rm QSL}(t^\ast)=B t^\ast + O(t^{\ast\,2})$ with 
\begin{equation}\label{eq:spectral2}
    B = \frac{\left|\frac{d}{dt}{\rm Tr}[\rho^2(t)]_{t=0}\right|}{2\nrm{\rho_0}_2\nrm{V}_2},
\end{equation}
which measures the magnitude of the initial purity-change rate normalized by the total Hilbert-Schmidt speed.
The Hilbert-Schmidt Cauchy-Schwarz inequality gives $0\le B \le 1$.  For the qubit case, we recover Eqs.~\eqref{eq:turn} and \eqref{eq:shortrp} with $K\to k$ and $B\to b$. We note that
Eq.~\eqref{eq:spectral} assumes a non-singular initial velocity which is not satisfied by the pure-state ensembles considered above; their finite-time indicators remain well defined, whereas their short-time TD expansion requires a separate treatment of the zero eigenspace.
In fact, we have not used Eqs.~\eqref{eq:spectral} and~\eqref{eq:spectral2} in the many-body examples presented here.

\emph{Conclusions.---}QSL constructions can provide distinct geometric probes of dissipative transients and QMEs. Their short-time limits reveal turning and alignment as essential features, rather than simply an initial relaxation speed.  Since the short-time QSL scores use information that typically pertains to times before the first crossing of trajectories, these geometric scores are able to anticipate Mpemba inversions before the actual crossings have occured.  The predictive power of short-time QSL scores for QMEs has been explicitly demonstrated for dissipative qubit and open many-body examples.  This framework identifies a concrete route toward predicting anomalous relaxation from limited dynamical information.  For many-body systems, where exact theoretical results can be too expensive, finding good indicators is of crucial importance.

\emph{Data availability.---}The data underlying the figures presented in this work will be made available at Zenodo. % \cite{Zenodo}.

\begin{acknowledgments} 
We acknowledge funding by the Deutsche Forschungsgemeinschaft (DFG, German Research Foundation) under Grant No.~277101999, TRR 183 (project B02), under Germany's Excellence Strategy - Cluster of Excellence Matter and Light for Quantum Computing (ML4Q) EXC 2004/2 - 390534769, and under Grant No.~EG 96/14-1, by the European Union under grant No. 101053159 (ERC - RAVE), and by a Google Quantum Research Award. 
\end{acknowledgments}

\bibliography{biblio}
%%%%%%%%%%%%%%%%%%%%%%%%%%%%%%%%%%%%%%%%%%%%%%%%%%%%%%%%%%%%%%%%%%%%%%%%%%%%%%%%%
% END MATTER
%%%%%%%%%%%%%%%%%%%%%%%%%%%%%%%%%%%%%%%%%%%%%%%%%%%%%%%%%%%%%%%%%%%%%%%%%%%%%%%%%
\newpage
\clearpage
\appendix
\section*{End Matter}
\setcounter{equation}{0}
\setcounter{figure}{0}
\renewcommand{\theequation}{A\arabic{equation}}
\renewcommand{\thefigure}{A\arabic{figure}}

We here present technical details and additional results. First,
using $\Delta\rho(t^\ast)=\rho(t^\ast)-\rho_0=\int_0^{t^\ast}\dot\rho(t)dt$, the triangle inequality gives
$
\nrm{\Delta\rho(t^\ast)}_1\leq\int_0^{t^\ast}\nrm{\dot\rho(t)}_1dt= \int_0^{t^\ast} \nrm{{\cal L}\rho(t)}_1 dt.
$
Similarly, applying also the Cauchy-Schwarz inequality,
$
|\Tr[\rho_0 \Delta\rho(t^\ast)]|\leq\nrm{\rho_0}_2\int_0^{t^\ast}\nrm{{\cal L}\rho(t)}_2dt.
$
These inequalities imply the QSL bounds \eqref{QSLbound}. Neither construction substitutes the unknown state at time $\tau$ for $\rho(t^\ast)$.  A stationary trajectory has a vanishing denominator
in Eqs.~\eqref{eq:qtd} and \eqref{eq:qrp}, and is removed from the eligible set \eqref{eq:eligible}. Note that the Hilbert-Schmidt norm need not be contractive for its integral inequality to hold.   

The TD-based QSL indicator \eqref{eq:qtd} discards radial amplitude information. To see this, consider 
the initial states  $\rho_{0,\tilde c}=\rss+\tilde c X$, with real $\tilde c>0$ and a Hermitian traceless operator $X$, where $\rho_{\tilde c}(t)=\rss+\tilde c e^{t{\cal L}} X$. Both the state distance between $\rho_{\tilde c}(t)$ and $\rho_{0,\tilde c}$ and the integrated trace-norm speed therefore scale with $\tilde c$. As a consequence, $\tau^{\rm QSL}_{\rm TD}(t^\ast)$ in Eq.~\eqref{eq:qtd} becomes independent of $\tilde c$. Using $D(t)=\frac{\tilde c}{2} \nrm{X(t)}_1$, however, the relaxation time $\tau_{\tilde c}=\inf\{t\geq0:D(t)\leq\epsilon\}$ generally depends on $\tilde c$. 
Radial invariance then limits relaxation-time ranking by TD-based QSLs, while same-ray states cannot
exhibit a Mpemba inversion. Other measures, e.g., the RP-based indicator in Eq.~\eqref{eq:qrp},
can provide complementary rankings for more general state families but are not needed to repair a classification failure in the present example.

Next we provide details concerning metrics and statistical analysis.
For given $\mathcal E_{t^\ast}$ in Eq.~\eqref{eq:eligible}, we define $N_0=|{\cal E}_{t^\ast}|$, the number of true positives $n_{\mathrm{TP}}=\#(M\cap\widehat M)$, of false positives $n_{\mathrm{FP}}=\#(\neg M\cap\widehat M)$, and of false negatives $n_{\mathrm{FN}}=\#(M\cap\neg\widehat M)$.
If a QSL indicator predicts a tie, it is not counted as Mpemba prediction.
For Eq.~\eqref{pi0q}, we then have 
$\pi_0=\frac{n_{\mathrm{TP}}+n_{\mathrm{FN}}}{N_0}$ and $q = \frac{n_{\mathrm{TP}}+n_{\mathrm{FP}}}{N_0}.$
The diagnostic quantities in Eq.~\eqref{eq:metrics} can be written as
$P=\frac{n_{\mathrm{TP}}}{n_{\mathrm{TP}}+n_{\mathrm{FN}}}$ and
$\FDR=\frac{n_{\mathrm{FP}}}{n_{\mathrm{TP}}+n_{\mathrm{FP}}}$.
The recall $P$ is undefined if there are no actual positives, while the FDR is undefined if there are no predicted positives. The PE is only defined if $\pi_0>0$ and $q>0$. 

We now turn to the short-time expansion for a dissipative qubit starting from Eq.~\eqref{affine}. For a smooth Bloch trajectory with $\mathbf j_0=\mathbf r^{(3)}(0)$ and $u_0=|\mathbf v_0|\ne 0$, we first obtain an analytical short-time expansion for the chord ($C$) and arc ($L$) lengths up to time $t^\ast$, which yields
 $L-C=u_0kt^{\ast\, 3}/24+O(t^{\ast\,4})$ and therefore $\tau_\td^{\rm QSL}(t^\ast)=t^\ast C/L$ gives Eq.~\eqref{eq:shorttd}. Both the initial velocity $u_0$ and the third-derivative term $\mathbf j_0$ cancel out from the bending correction.  For the RP expression, we use
$\nrm{\rho_0}_2=\sqrt{(1+|\mathbf r_0|^2)/2}, \, \nrm{\dot\rho}_2=|\dot{\mathbf r}|/\sqrt2,$ and
$
\Tr\left[\rho_0\dot\rho_0\right]=\frac12 \mathbf r_0\cdot\mathbf v_0.
$
As a result, we arrive at
Eq.~\eqref{eq:shortrp}. For $b=0$ or $k=0$, the next nonzero order must be retained.
However, higher orders may not exist; e.g., for $\rho_0\propto \mathds{1}$,  $\tau_{\rp}^{\rm QSL}(t^\ast)=0$ even though nonunital dynamics can evolve the state.

\begin{figure}[t]
\centering
\includegraphics[width=\linewidth]{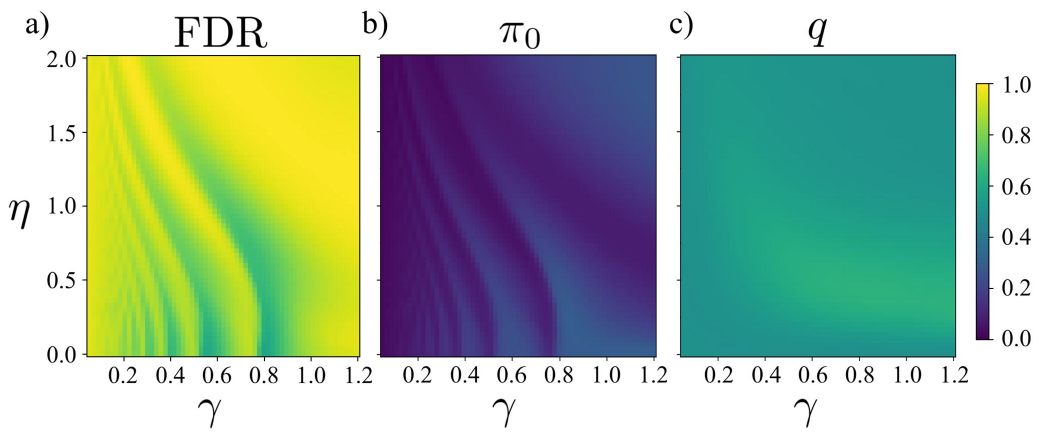}
\caption{Color-scale plots in the $\gamma$--$\eta$ plane of the planar qubit model in Eq.~\eqref{eq:A}
for (a) FDR, (b)  $\pi_0$, and (c)  $q$, see Eqs.~\eqref{pi0q} and \eqref{eq:metrics}; see Fig.~\ref{fig2} for $P$ and PE. We use the short-time QSL score $s_{\rm TD}=-k$, see Eq.~\eqref{kdef}, and $\epsilon=10^{-2}$, averaging over initial-state pairs as described in the text. 
}  
\label{figS1}
\end{figure}

\begin{figure}[t]
\centering
\includegraphics[width=0.71\linewidth]{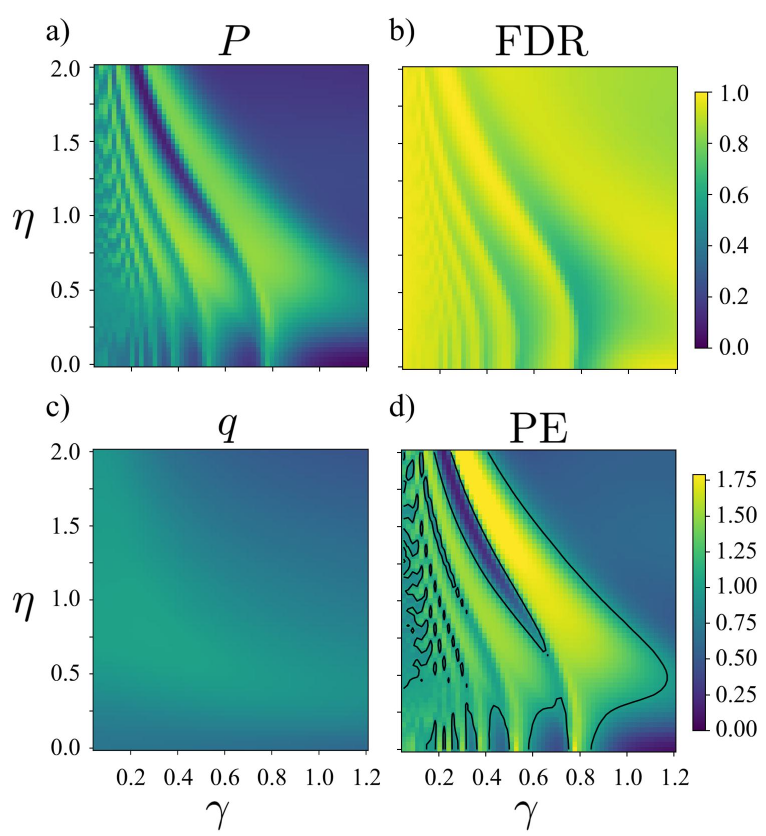}
\caption{Color-scale plots of (a) $P$, (b) FDR, (c) $q$, and (d) PE in the $\gamma$--$\eta$ plane for the qubit model as in Fig.~\ref{fig2} and Fig.~\ref{figS1}, but using the QSL score $s_{\rm RP}=b$ in Eq.~\eqref{eq:shortrp}. Here, $P$, FDR, and $q$ share the same $[0,1]$ color-scale bar.
Note that $\pi_0$ does not depend on the QSL score, see Fig.~\ref{figS1}(b).  
} 
\label{figS2}
\end{figure}

We next turn to the exactly solvable planar model for a dissipative qubit described by the affine equation \eqref{affine} for the Bloch vector in the $xy$ plane with Eq.~\eqref{eq:A}. 
Writing ${\bf v}_0=|{\bf v}_0| (\cos\phi,\sin\phi)^T$, we find for $k$ in Eq.~\eqref{eq:turn} the analytical result
\begin{equation}\label{kdef}
k=4[1+\gamma\cos(2\phi)]^2.
\end{equation}
To obtain the exact propagator,  we set $\Lambda_\perp=-2\gamma(1+\eta^2)\mathds{1}+B$ with 
$
B=\begin{pmatrix}
0&2(\gamma-1)\\
2(\gamma+1)&0
\end{pmatrix},
$
where $B^2=4(\gamma^2-1)\mathds{1}$.
For $\gamma<1$, let $\Omega=2\sqrt{1-\gamma^2}$. Then
\begin{equation}\label{eq:eL}
e^{\Lambda_{\perp}t}
=e^{-2\gamma(1+\eta^2)t}
\left[\mathds{1}\cos(\Omega t)
+\frac{B}{\Omega}\sin(\Omega t)
\right].
\end{equation}
For $\gamma>1$, one replaces $\Omega\to\chi=2\sqrt{\gamma^2-1}$ and $(\cos,\sin)\to (\cosh,\sinh)$. At the exceptional point $\gamma=1$, the bracket $[\ldots]$ in Eq.~\eqref{eq:eL} is replaced by $\mathds{1}+tB$. The stationary Bloch vector is
$
\mathbf r_{\rm st}
=\frac{2\gamma\eta}
{1+\gamma^2\eta^2(2+\eta^2)}
\begin{pmatrix}
1+\gamma\eta^2\\
1-\gamma\eta^2
\end{pmatrix},
$
and $D(t)=\frac12 |e^{\Lambda_{\perp}t}(\mathbf r_0-\mathbf r_{\rm st})|.$
One then obtains $\tau$ without having to diagonalize ${\cal L}$. We note that $D(t)$ is always monotonic as required by contractivity. 
The planar eigenvalues $\lambda_\pm$ specified in the main text
 reveal a transition from oscillatory to overdamped
relaxation as $\gamma$ increases through $\gamma=1$.
For $\gamma<1$, we have $k>0$ for all $|{\bf v}_0|\ne 0$, see Eq.~\eqref{kdef}.
For $\gamma>1$, however, along the single-mode trajectories
$\mathbf r(t)=\mathbf r_{\mathrm{st}}+c_\pm e^{\lambda_\pm t}\mathbf e_\pm$, where $\mathbf{e}_\pm$ are the  eigenvectors, velocity and acceleration are parallel, implying $k=0$ in Eq.~\eqref{kdef}.
(For $\gamma=1$, only one such eigenvector exists.) In these cases, 
the TD-QSL  gives simply $\tau^{\rm QSL}_\td(t^\ast)=t^\ast$ and therefore 
fails to predict a QME. For  generic initial states, however, both eigenmodes contribute,
producing a curved trajectory; in general, one then finds $k> 0$. This identifies a mechanism by which the TD score can lose discriminatory power in the overdamped regime, see Fig.~\ref{fig2} and 
Fig.~\ref{figS1}.

\begin{figure}[t]
\includegraphics[width=\linewidth]{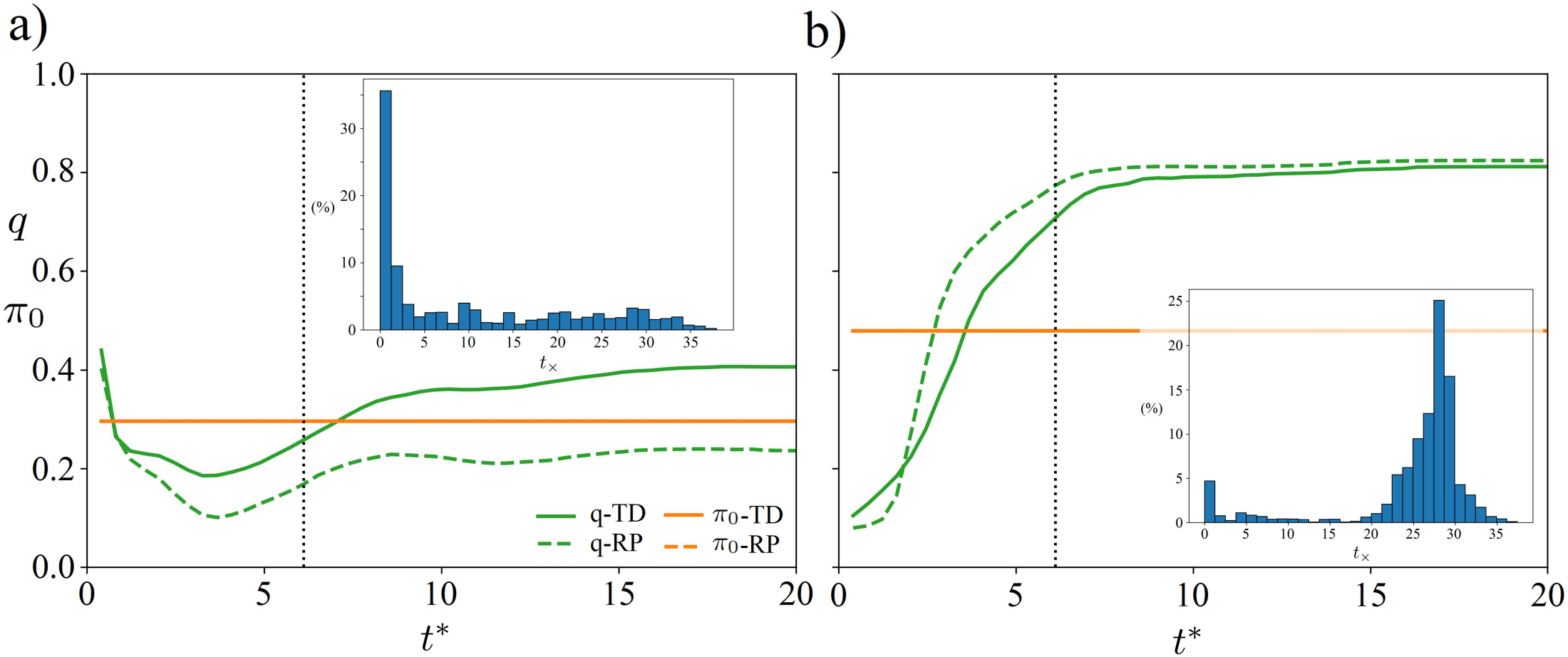}
\caption{$\pi_0$ and $q$ vs $t^\ast$ 
for the collective-spin model \eqref{eq:collective} with the parameters in Fig.~\ref{fig3} 
and finite-time TD-based (solid) or RP-based (dashed curves) QSL indicators.  
Vertical dotted lines indicate the inverse Liouvillian gap.
(a) Results for the initial states $|\psi_1\rangle$ in Eq.~\eqref{eq:three_dicke_family}, using the same averaging 
procedure as in Fig.~\ref{fig3}.  Inset: Histogram of Mpemba pairs vs the last crossing time $t_\times$ after averaging over initial state pairs.
(b) Same as in (a) but for the initial states $|\psi_2\rangle$ in Eq.~\eqref{eq:dicke_initial_state}.}
\label{figS3}
\end{figure}

In Fig.~\ref{fig2}, we have presented results for $P$ and $\mathrm{PE}$, see Eqs.~\eqref{eq:metrics} and \eqref{precisionenrichment}, using the exact solution and  $s_\td=-k$.
In Fig.~\ref{figS1}, we show  numerical results for FDR, $\pi_0$, and $q$.  We have sampled over $N_s=600$ initial states quasi-uniformly distributed over the $xy$ Bloch disk using a deterministic golden-angle construction. Specifically, $N_s/2$ points are generated with polar coordinates $r_j=\sqrt{(2j+1)/N_s}$ and $\phi_j=\phi_0+j\pi(3-\sqrt{5})$, where $j=0,\ldots,N_s/2-1$. The radial spacing ensures uniform coverage in area, while the golden-angle increment minimizes angular clustering. The remaining $N_s/2$ points are obtained by inversion, $(x_j,y_j)\to(-x_j,-y_j)$, ensuring  antipodal symmetry. This set of initial states is used for all Bloch-disk scans.  The FDR is rather large because of the small $\pi_0$.  We also observe that while the predictions of TD vs RP scores differ in detail, they agree in overall features.
Indeed, in Fig.~\ref{figS2}, we show results obtained from $s_\rp=b$.  The parameter regions where 
QMEs are predicted with ${\rm PE}>1$ and large recall $P$ are similar for both scores. For weak damping, both $k$ and $b$ can be rather large. In such cases, short-time geometry encoded in QSL scores can well predict 
Mpemba orderings.

\begin{figure}[t]
\includegraphics[width=\linewidth]{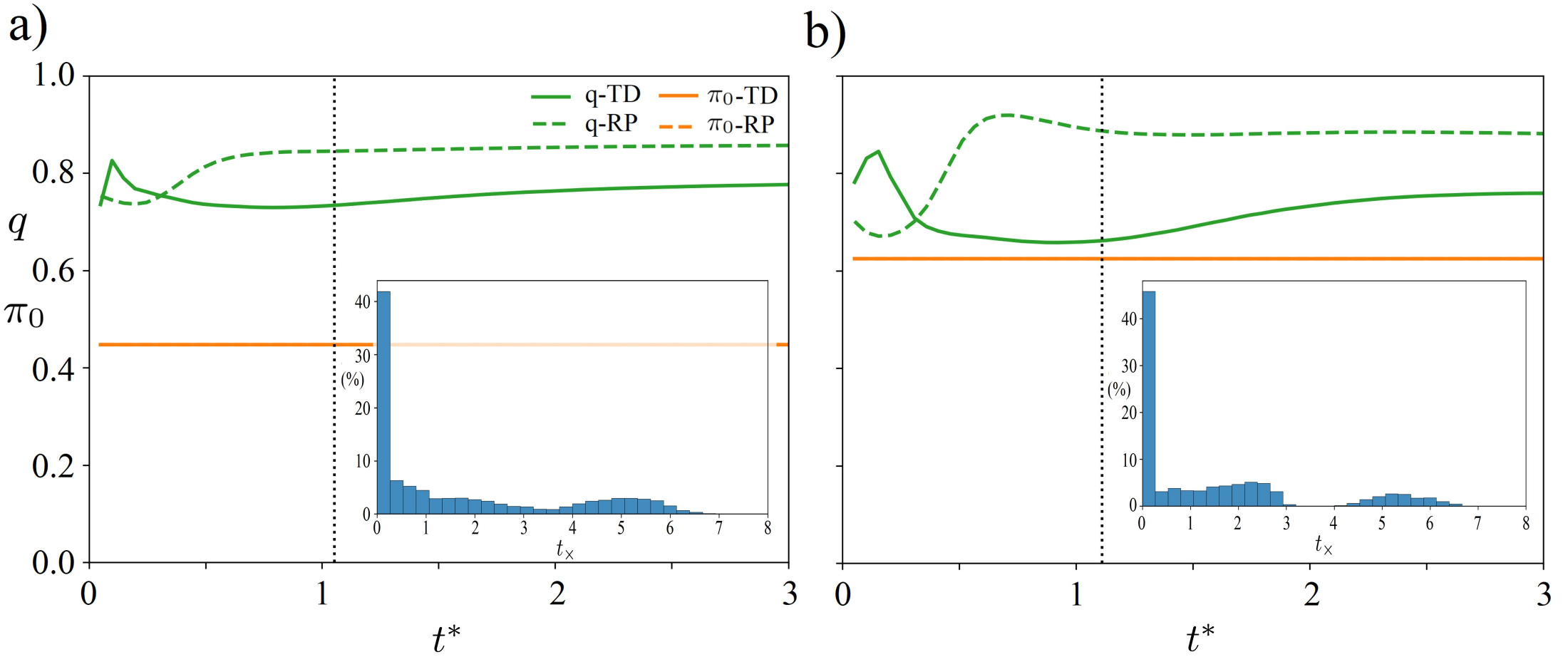}
\caption{$\pi_0$ and $q$ vs $t^\ast$ for the open spin chain, see Eq.~\eqref{eq:chain},
with the parameters in Fig.~\ref{fig4}, using TD-based (solid) and RP-based (dashed curves) QSL indicators.
Vertical dotted lines indicate the inverse Liouvillian gap. The same averaging procedure as in Fig.~\ref{fig4} was used.
(a) Interacting case with parameters as in Fig.~\ref{fig4}(a). The inset shows a histogram of Mpemba pairs 
vs $t_\times$. 
(b) Same as (a) but for the noninteracting case, with parameters as in Fig.~\ref{fig4}(b).   }
\label{figS4}
\end{figure}
 
We now discuss additional finite-$t^\ast$ QSL indicator results, see Eqs.~\eqref{eq:qtd} and \eqref{eq:qrp}, for the many-body models in the main text.  First, in Fig.~\ref{figS3}, we show the $t^\ast$-dependence of $\pi_0$ and $q$, see Eq.~\eqref{pi0q}, for the collective-spin model \eqref{eq:collective}, using the same parameters as in Fig.~\ref{fig3}. The two panels in Fig.~\ref{figS3} are again for the two initial-state classes studied in Fig.~\ref{fig3}, where the respective insets show histograms for the percentage of Mpemba pairs whose last crossing time
is $t_\times$ after averaging over initial-state pairs.  In Fig.~\ref{figS3}(a), corresponding to the initial states $|\psi_1\rangle$ in Eq.~\eqref{eq:three_dicke_family}, the inset shows a dominance of early last crossing times. In such cases, short-time QSL indicators are especially useful and accurate.  The main panel shows $\pi_0$, which does not depend on the QSL indicator, and $q$, which has comparable values for both QSL indicators.  In Fig.~\ref{figS3}(b), where we take the initial-state class $|\psi_2\rangle$ in Eq.~\eqref{eq:dicke_initial_state}, last crossing times tend to prevail at late times.  
The $t^\ast$-dependence of $\pi_0$ comes only from changes in the eligible population, while $q(t^\ast)$ 
recovers its plateau value at times beyond the inverse Liouvillian gap.
In Fig.~\ref{figS4}, we address the open spin-chain model, see Eq.~\eqref{eq:chain} and Fig.~\ref{fig4}, and show the
corresponding results for $\pi_0(t^\ast)$ and $q(t^\ast)$ as well as histograms for the distribution of $t_\times$.  The two panels correspond to the two panels in Fig.~\ref{fig4}, showing an interacting vs noninteracting case.
In both cases, a dominance of early times $t_\times$ is observed, where $q(t^\ast)$ 
rapidly approaches its plateau value even before the inverse Liouvillian gap has been reached.

%We finally address slow sectors and the crossing criterion for degenerate or complex-valued slow eigenvalues.
%For a real isolated slow eigenvalue $\lambda=-g$ with nonzero amplitude, $D(t)= C e^{-gt}[1+o(1)]$ implies $\tau(\epsilon)= g^{-1}\log(C/\epsilon)+o(1)$, which in turn implies a threshold-independent ordering for unequal initial-state amplitudes for $\epsilon\to 0$. 
%A degenerate real slow eigenvalue instead requires a complete projection since suppressing only one coefficient does not remove the entire sector \cite{Benjadi2026}; near-degenerate modes can also coexist for long intervals \cite{Macieszczak_2016}. Complex slow pairs and defective blocks require a separate treatment \cite{Benjadi2026}. We expect that the resulting spectral structures are not determined by the initial turning scores alone.

\end{document}